\documentclass[aps,nofootinbib,superscriptaddress,twocolumn,prd,10pt]{revtex4-1}

\usepackage{graphicx}
\usepackage{amsmath,amssymb}
\usepackage{hyperref}
\usepackage{braket}
\usepackage{booktabs}
\usepackage{subfigure}
\usepackage{float}
\usepackage[dvipsnames]{xcolor}
\usepackage{mathrsfs}
\usepackage{comment}
\newcommand{\DE}{{\rm DE}}
\hypersetup{
    colorlinks=true,
    linkcolor=red,
    citecolor=blue,
}

\begin{document}

\title{Can Late Dark Energy Transitions Raise the Hubble constant?}

\author{Giampaolo Benevento}
\affiliation{Dipartimento di Fisica e Astronomia ``G. Galilei'', Universit\`a degli Studi di Padova, via Marzolo 8, I-35131, Padova, Italy}\affiliation{
INFN, Sezione di Padova, via Marzolo 8, I-35131, Padova, Italy}

\author{Wayne Hu}
\affiliation{Kavli Institute for Cosmological Physics, Department of Astronomy \& Astrophysics, Enrico Fermi Institute, The University of Chicago, Chicago, IL 60637, USA}

\author{Marco Raveri}
\affiliation{Center for Particle Cosmology, Department of Physics and Astronomy, University of Pennsylvania, Philadelphia, PA 19104, USA}

\begin{abstract}
Late times dark energy transitions at redshifts $z \ll 0.1$ can raise the predicted value of the Hubble constant to the SH0ES value, $74.03\pm 1.42$ (km\,s$^{-1}$\,Mpc$^{-1})$ or more, while providing an equally good fit as $\Lambda$CDM at $67.73 \pm 0.41$ 
to higher redshift data, in particular from the cosmic microwave background and baryon acoustic oscillations.   
These models however do not fully resolve the true source of tension between the distance ladder and high redshift observations: the local calibration of supernovae luminosities well out into the Hubble flow.  
When tested in this manner by transferring the SH0ES calibration to the Pantheon supernovae dataset, the ability of such transitions to raise the Hubble constant is reduced to $69.17 \pm 1.09$. 
Such an analysis should also be used when testing any dynamical dark energy model which can produce similarly fine features in redshift or local void models.
\end{abstract}

\maketitle
\section{Introduction}
The statistically significant disagreement within the $\Lambda$CDM model between the value of the Hubble constant measured by the local distance ladder and that inferred from measurement of cosmological observables
anchored at recombination by the cosmic microwave background (CMB) is currently the strongest challenge to $\Lambda$CDM as the standard model of cosmology.

In particular in $\Lambda$CDM, the Planck measurement of $H_0=67.4 \pm 0.5$ (in units of km s$^{-1}$ Mpc$^{-1}$ assumed throughout)~\cite{Aghanim:2018eyx} is in $4.4\sigma$ tension with the latest SH0ES estimate~\cite{Riess:2019cxk} $H_0=74.03 \pm 1.42$  based on Type Ia Supernovae (SN) in the Hubble flow. 
This measurement of $H_0$ requires a calibration of the peak luminosity of SN using the local distance ladder. 
The most precise and mature method, adopted by the SH0ES collaboration, consists in using  Cepheid variables as intermediate calibrators, but other approaches are also possible. 
Using instead a calibration based 
on the tip of the red giant branch, the Chicago Carnegie Hubble Project~\cite{Freedman:2019jwv} finds  $H_0=69.6 \pm 0.8 ({\rm stat}) \pm 1.7 ({\rm syst}) $ \cite{Freedman:2020dne}, though its robustness to systematic errors has been debated~\cite{yuan2019consistent}. A third approach uses Mira variables to calibrate
SN and currently gives $H_0=72.7 \pm 4.6$ \cite{2020ApJ...889....5H}.
On the other hand using a completely independent method of time-delays of multiply imaged quasars, the H0LiCOW collaboration finds $H_0= 73.3^{+1.7}_{-1.8}$~\cite{Wong:2019kwg}, while in complete agreement with the SN on the amplitude and shape of the distance-redshift relation~\cite{Pandey:2019yic}.    
Likewise, the Megamaser Cosmology Project provides independent geometric measurements of $H_0=73.9\pm 3$ from accretion disks around active galactic nuclei of galaxies in the Hubble flow \cite{Pesce:2020xfe}.

As emphasized by Refs.~\cite{Bernal:2016gxb, Aylor:2018drw, Knox:2019rjx}, the only viable single point solutions to this tension are those that change either the anchor at recombination or for local measurements because of the wealth of intermediate measurements that connect them.  
Whereas local explanations mainly invoke astrophysical and systematic uncertainties in the local calibration, much recent attention has been focused on adding a so-called ``early" dark component which is significant only near recombination to alter the CMB anchor in a specific way~\cite{Poulin:2018cxd,Agrawal:2019lmo,Lin:2019qug}.   

In this paper, we consider the converse where there is a late transition in the dark energy density only very near the present epoch, $z< 0.1$.  
Such a transition would escape detection from observables at much higher redshift~\cite{Mortonson_2009}.   
These models also highlight the difference between raising the value of the Hubble constant, in a manner compatible with the CMB and other
high redshift observables, and truly solving the problem underlying the Hubble tension.   
Since the local distance ladder calibrates SN far into the Hubble flow, if this transition occurs too recently it may raise $H_0$ without actually changing the part of the Hubble diagram where the tension is inferred.    

In such cases, which we dub late dark energy (LDE) transition models, the traditional analysis of using the SH0ES measurement as a constraint on $H_0$  is misleading.  
A proper analysis requires considering the local distance ladder information as calibrating the absolute magnitude of higher redshift
SN.   
We adopt the approach introduced in Ref.~\cite{raveri2019quantifying} where the SH0ES calibration is transferred to the Pantheon SN dataset~\cite{Scolnic:2017caz} and contrast results with the traditional approach. 

In \S \ref{Sec:LDEtheory}, we review the LDE transition scenario itself \cite{Mortonson_2009}.
We discuss the methodology for analyzing the local distance ladder information as a measurement of $H_0$ vs.~a calibration of SN and describe our baseline high redshift datasets in \S \ref{Sec:Datasets}.    
We present results in \S \ref{Sec:Results} and discuss them in 
\S \ref{Sec:Outlook}.

\section{Late Dark Energy} \label{Sec:LDEtheory}
Following Ref.~\cite{Mortonson_2009} we consider a late dark energy modification of the $\Lambda$CDM expansion history leading to a fractional change of $\delta$ in the Hubble constant
\begin{equation} \label{eq:H0}
H_0^2 = \tilde{H}_0^2 (1+ 2 \delta) 
\end{equation}
from $\tilde{H}_0$, the prediction for a flat $\Lambda$CDM model with a cosmological constant density $\tilde{\rho}_{\Lambda}$.
This can be obtained in a dark energy scenario 
in which the cosmological constant density $\tilde{\rho}_{\Lambda}$ is modulated by a smooth step function $f(z)$,
\begin{equation}
\rho_{\DE}(z)= \left[1+f(z)\right] \tilde{\rho}_{\Lambda} , \label{eq:rho_DE}
\end{equation}
where 
\begin{flalign}
&f(z) = \frac{2 \delta}{\tilde{\Omega}_{\Lambda}}\frac{S(z)}{S(0)} \ , \\
&S(z)= \frac{1}{2} \left[ 1- \tanh \left(\frac{z-z_{t}}{\Delta z} \right) \right],
\end{flalign}
and $\Delta z$ is the duration of the transition.  For $z\gg z_t$, the expansion history is
thus indistinguishable from the reference  $\Lambda$CDM model.  We shall see that this property allows
the LDE model to mimic all of the high redshift observables of the reference $\Lambda$CDM model.
The LDE model we consider is therefore completely specified by the choice of the three parameters \{$\delta$, $z_t$, $\Delta z$\} in addition to the standard $\Lambda$CDM parameters which control the $z\gg z_t$ universe: $\theta_{\rm MC}$, the effective angle subtended by the CMB sound horizon at recombination; $\Omega_b h^2$, the physical baryon density; $\Omega_c h^2$, the physical cold dark matter density, $\tau$, the Thomson optical depth to recombination;  $A_s$, the normalization of the curvature power spectrum at $k=0.05\,$Mpc$^{-1}$; and $n_s$, its tilt.  $H_0$ itself is derived from these model parameters.

Although not necessary for our analysis, we can  relate the LDE model to a physical model for the dark energy
through its equation of state
\begin{equation}
1+w= \frac{1}{3} \frac{(1+ z)f'}{1+f} .
\end{equation}
As shown in Ref.~\cite{Mortonson_2009}, this equation of state can be achieved with a minimally coupled scalar field $\phi$ with the potential
\begin{flalign}
V(\phi(z)) &= \frac{1}{2}(1+ w(z))\rho_{\DE}(z) \nonumber  \\
     &= [(1+f) - (1+z) f'/6] \tilde{\rho}_{\Lambda}.
\end{flalign}
A positive value of $\delta$, i.e.\ $H_0> \tilde{H}_0$, requires a phantom equation of state $w<-1$. The sign of the kinetic term must therefore change with $\delta$, leading to a scalar field Lagrangian 
\begin{equation}
L =\frac{{\rm sgn}(\delta)}{2} g^{\mu\nu} \partial_\mu\phi \partial_\nu\phi - V(\phi).
\end{equation}
While $\delta>0$ implies a ghost, which is unstable at the quantum level, our aim is to illustrate the phenomenology of a model with this $w(z)$ at the classical level where no instability arises.
 
We compute observables associated with the LDE model using the Boltzmann solver EFTCAMB~\cite{Hu:2013twa,Raveri:2014cka,hu2014eftcambeftcosmomc}, a modification of the CAMB~\cite{Lewis:1999bs} code.
EFTCAMB uses the  effective field theory of dark energy formalism \cite{2013JCAP...02..032G, 2013JCAP...08..025G} where 
\begin{equation}
c(z) = \frac{1}{2} \rho_{\DE} (1+ w(z)), \ \ \ \Lambda(z)=  \rho_{\DE} w(z),
\end{equation} 
give the mapping from LDE model parameters to EFTCAMB functions $\{ c(z),\Lambda(z) \}$.

\section{Datasets and methodology} \label{Sec:Datasets}
We test the LDE model predictions against several complementary datasets that are in tension with the local distance ladder under $\Lambda$CDM. 
For the exploration of the parameter posterior distribution, we use the Markov Chain Monte Carlo (MCMC) algorithm, implemented in the CosmoMC~\cite{Lewis:2002ah} code.
For the statistical analysis of the posterior distributions, we employ the GetDist code~\cite{Lewis:2019xzd}.

For the high-$z$ side of the Hubble tension, we use the Planck 2018 measurements of CMB temperature, polarization, and lensing power spectra 
(multipoles range $8\leq \ell \leq 400$)~\cite{Aghanim:2018eyx, Aghanim:2019ame}.
We further include baryon acoustic oscillations (BAO) measurements of BOSS galaxies in its DR12 data release~\cite{Alam:2016hwk}, the Sloan Digital Sky Survey (SDSS) main galaxy sample~\cite{Ross:2014qpa}, and the 6dFGS survey~\cite{Beutler:2011hx}.
We employ all of the standard nuisance parameters and $\Lambda$CDM priors in these analyses~\cite{Aghanim:2019ame}.   

Central to our analysis is the  Pantheon Supernovae sample, which combines Supernova Legacy Survey, SDSS and HUbble Space Telescope supernovae with several low redshift ones~\cite{Scolnic:2017caz}, spanning the redshift range $z \in [0.01, 2.26]$.
The publicly available release of the Pantheon catalog provides SN magnitudes corrected for systematics effects, such as the stretch of the light curve, the color at maximum brightness, and the stellar mass of the host galaxy.  The apparent magnitude $m$ of each SN after correction is referenced to an arbitrary fiducial absolute magnitude $M_{\mathrm{fid}}=-19.34$
that corresponds to a fiducial value of the Hubble constant of $H_0^{\rm fid}=70$  in the $\Lambda$CDM model.\footnote{\texttt{\url{https://github.com/dscolnic/Pantheon}} provides $m-M_{\rm fid}$ in the $H_0^{\rm fid}=70$ convention of~\cite{Guy:2005me}.}
The likelihood of the data for a given luminosity distance $d_L$ and true fiducial absolute magnitude $M$ is
\begin{equation} 
{\cal L}_{\rm SN} = {\cal N} (m - M_{\rm fid}; 5 \log_{10} \frac{d_L}{10 {\rm pc}} + M-M_{\rm fid} , \Sigma) ,
\label{eq:LSN}
\end{equation} 
where ${\cal N}(x,\bar x, \Sigma)$ denotes a normal distribution for a data vector $x$, with mean $\bar x$,
and covariance $\Sigma$.   The unknown parameter $M$ is then marginalized.   
We call the combination of Planck, BAO, and Pantheon data the baseline dataset.

In order to test the impact of late transitions in the expansion history on the Hubble tension, we also consider the local calibration of the distance ladder from SH0ES~\cite{Riess:2019cxk}.
This has been included in our analysis, following two different approaches. 
In the first approach, following what is usually done in similar analyses of dynamical dark energy, the SH0ES constraint is implemented at redshift $z=0$ as $H_0=74.03\pm 1.42$ and  added to our baseline dataset. 
We denote this configuration as baseline$+H_0$.

This approach suffers from the fact that if the transition, or more generally the dark energy equation of state, varies substantially between the calibrators of the distance ladder and the SN in the Hubble flow that are used by SH0ES, which goes out to $z \sim 0.15$, fitting $H_0$ does not necessarily mean fitting the SH0ES data or resolving the actual origin of the $H_0$ tension. 

Therefore, in an alternative approach, adopted by~\cite{raveri2019quantifying, p2019model, Camarena_2020}, the local distance ladder is not implemented as a separate constraint on $H_0$ but is used to calibrate the Pantheon SN fiducial absolute magnitude $M$ in Eq.~(\ref{eq:LSN}).
The absolute magnitude and its standard deviation $\sigma_{M}$ are related to $H_0$ and $\sigma_{H_0}$ in $\Lambda$CDM by
\begin{align}
\bar M &= 5 \log_{10} \frac{H_0}{H_0^{\mathrm{fid}}} +M_{\mathrm{fid} }\approx  -19.22 \,, \\
\sigma_{M} & = \frac{5}{\ln 10} \frac{\sigma_{H_0}}{H_0} \approx 0.042 \,.
\label{eq:Merror}
\end{align}
The modified version of the Pantheon likelihood + SH0ES calibration is then
\begin{equation} 
{\cal L} = {\cal L}_{\rm SN} \times {\cal N} (M,\bar M, \sigma_M^2).
\end{equation}
When combined with the Planck and BAO data, we call this dataset baseline$+ M$. 
 
The two different treatments lead to completely equivalent results for the $\Lambda$CDM model but entirely different results for the LDE model as we shall see.

\begin{table*}[!ht]
\setlength{\tabcolsep}{12pt}
\centering
\begin{tabular}{@{}cccccc@{}}
\colrule                                                                                                               
$\Lambda$CDM                & baseline          &  baseline+$H_0$        \\
\colrule
$100\theta_{\rm MC}$   & 1.04110 ($1.04110\pm0.00028$)   & 1.04120 ($1.04118 \pm 0.00029$)  \\
$\Omega_bh^2$          & 0.02243 ($0.02243\pm0.00013$)   & 0.02258 ($0.02253\pm0.00013$)       \\
$\Omega_ch^2$          & 0.11911 ($0.11917\pm0.00092$)      & 0.11792 ($0.11817\pm 0.00086$)       \\
$\tau$                 & 0.0560 ($0.0565\pm0.0073$)         & 0.064 ($0.0595\pm0.0076$)          \\
$\ln(10^{10}A_s)$      & 3.047 ($3.047\pm0.014$)        & 3.059 ($3.051\pm0.015$)         \\
$n_s$                  & 0.9677 ($0.9671\pm0.0036$)      & 0.9714 ($0.9696\pm0.0036$)   \\
\colrule
$H_0$                  &  67.75  ($67.73\pm0.41$)          &  68.36 ($68.22\pm0.39$)        \\
\colrule
$\Delta\chi^2_{\rm base}$ & 0                            & 2.2 \\
$\Delta\chi^2_{{\rm base}+H_0} $ & 0                            & $-1.2$ \\
$\Delta\chi^2_{{\rm base}+M} $ & 0                            & $-1.2$ \\
\end{tabular}
\caption{\label{tab:LCDM} $\Lambda$CDM maximum likelihood (ML), mean and 1$\sigma$ uncertainties for parameters under the baseline vs.\ baseline+$H_0$ dataset analyses.    The former, referred to as $\Lambda$CDM ML,
provides the reference model against which we quote both the
$\Delta\chi^2_{\rm base}$ between models for the baseline data as well as
for the baseline+$H_0$ and baseline+$M$ datasets, 
$\Delta\chi^2_{{\rm base}+H_0}$ and $\Delta\chi^2_{{\rm base}+M}$, here and throughout.  
The $\Lambda$CDM+$H_0$ ML compromises the fit to the baseline data by $2.2$ in order to better fit the $H_0$ data.   This model performs equally well in   $\Delta\chi^2_{{\rm base}+M}$  and
likewise the baseline+$M$ ML and analysis  (not shown) give results that are indistinguishable.}
\end{table*}

\section{Hubble Tension vs $H_0$ Values} \label{Sec:Results}

We begin with the $\Lambda$CDM model to establish the baseline and test our two ways of treating the local distance ladder as a constraint on $H_0$ vs\ on the absolute magnitude of SN $M$ deep in the Hubble flow.   In Tab.~\ref{tab:LCDM} (middle column), we first show the $\Lambda$CDM constraints and maximum likelihood (ML)
model using only the baseline data, recovering $H_0=67.75$ for the latter as expected.   When testing this model against either the $H_0$ or the $M$ representation of the distance ladder data, this model is a bad fit, but establishes the baseline against
which improvements can be measured.    
A good fit that resolves the Hubble tension should therefore have 
\begin{align}
\Delta\chi^2_{{\rm base}+H_0}  & \sim -(4.4)^2 \sim -19 \nonumber\\
& \sim \Delta\chi^2_{{\rm base}+M},
\label{eqn:resolution}
\end{align} 
relative to this model.  We therefore use this $\Lambda$CDM ML model as the standard for comparison throughout for $\chi^2$ values and model differences with Eq.~(\ref{eqn:resolution}) as the criteria for resolving the Hubble tension.

To clarify this approach, we also analyze the $\Lambda$CDM model against the baseline $+H_0$ or $+M$ data.   
The two approaches in $\Lambda$CDM are completely equivalent and we list $+H_0$ in Tab.~\ref{tab:LCDM}.  Notice that the maximum likelihood model now compromises and sets $H_0=68.36$, providing a worse fit to the baseline dataset $\Delta\chi^2_{\rm base}=2.2$,
but a better fit to the total data $\Delta\chi^2_{{\rm base}+H_0}=-1.2$ relative to  $\Lambda$CDM ML model.
For brevity, we call this alternate model the $\Lambda$CDM+$H_0$ ML model and use this naming convention for other
models below. This model performs equally well in $\Delta\chi^2_{{\rm base}+M}$, but
neither satisfy Eq.~(\ref{eqn:resolution}) for Hubble tension resolution.
We also obtain nearly identical results for the $\Lambda$CDM$+M$  ML model and parameter constraints
and so do not list them explicitly in Tab.~\ref{tab:LCDM}.

\begin{table*}[th]
\setlength{\tabcolsep}{12pt}
\centering
\begin{tabular}{@{}cccccc@{}}
\colrule
LDE             & baseline+$H_0$       & baseline+$M$             \\
\colrule                                                                                                               
$100\theta_{\rm MC}$    & 1.04101 ($1.04104\pm0.00029$) & 0.104113 ($1.04114\pm 0.00029$) \\
$\Omega_bh^2$          & 0.02247 ($0.02244\pm 0.00013$) & 0.02253 ($0.02251\pm 0.00014$)        \\
$\Omega_ch^2$              & 0.11905 ($0.11916\pm 0.00091$) & 0.11843 ($0.11851\pm 0.00091$)    \\
$\tau$                  & 0.0561 ($0.0567^{+0.0067}_{-0.0074}$)   & 0.0590 ($0.0585^{+0.0070}_{-0.0078}$)     \\
$\ln(10^{10}A_s)$      & 3.045 ($3.048\pm 0.014$)       & 3.051 ($3.050 \pm 0.015$)      \\
$n_s$              &  0.9688 ($0.9672\pm 0.0036$)    & 0.9703 ($0.9688\pm 0.0037$) \\
\colrule                                                                                                               
$\delta$      & 0.096 (0.074$\pm$ 0.030)        & 0.015 ($0.017^{+0.014}_{-0.016}$)        \\
$\Delta z$     & 0.0015 ($< 0.035$)         & 0.0059 (prior limited) \\
$z_t$                      & 0 ($< 0.027 $)               & 0.043 (prior limited)  \\
\colrule
$H_0$                     &  74.01 (72.$5\pm$ 1.85)        & 69.06 (69.17 $\pm$ 1.09) \\
$M$                     & --
& -19.396 (-19.399  $\pm$ 0.012) \\
\colrule                                                                                                               
$\Delta\chi^2_{\rm base}$   & 0   & 0.9                     \\
$\Delta\chi^2_{{\rm base}+H_0}$ & $-19.5$                           & $-6.3$                     \\
$\Delta\chi^2_{{\rm base}+M}$ & $-0.75$                           & $ -3.0$                     \\
\end{tabular}
\caption{ \label{tab:parameters}
LDE maximum likelihood (ML) parameters and  constraints as in Tab.~\ref{tab:LCDM}.  
Local distance ladder data are added to the baseline data, either as a direct constraint on the $z=0$ expansion rate ($+H_0$) or as a constraint on the absolute magnitude of Pantheon SN ($+M$).
$\Delta\chi^2$ values are relative to the $\Lambda$CDM ML model in Tab.~\ref{tab:LCDM} (center column).  
Upper limits are quoted at $95\%$CL. 
}
\end{table*}

\begin{figure}[t]
\centering
\includegraphics[width=\columnwidth]{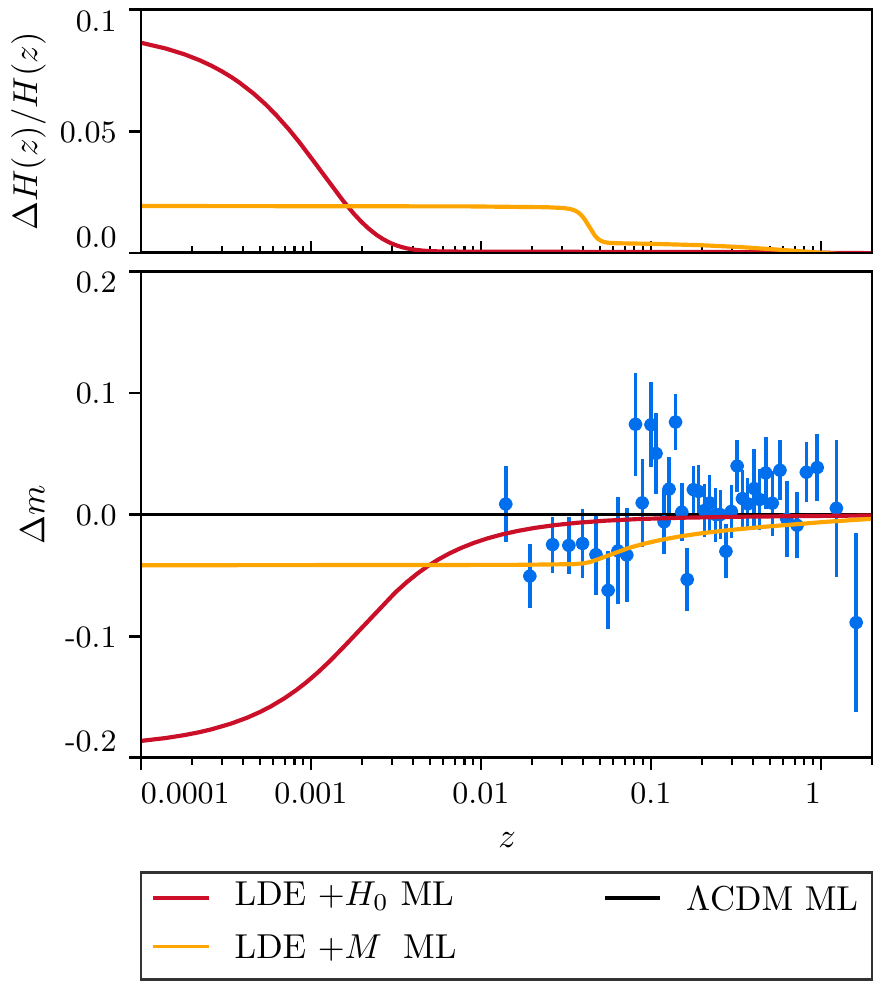}
\caption{
\label{fig:SN_residual}
Top: the deviation in the Hubble rate $H(z)$ for the LDE+$H_0$ and LDE+$M$  ML models (Tab.~\ref{tab:parameters}) relative to the baseline $\Lambda$CDM ML model (Tab.~\ref{tab:LCDM}).  Bottom: Pantheon SN magnitude residuals for the
same, offset from the SH0ES calibration by $\Delta M = 0.192$ (see Eq.~\ref{eq:Mshift}) to emphasize the shape.   Both models
fit the shape of the SN residuals, LDE+$H_0$ has $H_0\approx 74$,  yet both fail
to account for this large offset. }
\end{figure}

For the LDE model, we obtain very different results for the $+H_0$ vs $+M$ cases given in Tab.~\ref{tab:parameters}.
In order to study late time transitions, we place a flat prior of $0\le z_t \le 0.05$ and $0\le \Delta z \le 0.05$ while leaving the flat prior on $\delta$ uninformative.  
For the $+H_0$ case, the data favor an upward transition for the expansion rate of $\delta = 0.096$ (or $H_0\approx 74$).
The  LDE+$H_0$ ML model pins $z_t$ to the lowest redshift allowed, with a finite but small enough width $\Delta z\approx 0.0015$ so that
the transition occurs but is confined below the lowest SN redshift. 

In Fig.~\ref{fig:SN_residual}, we show the fractional change in the Hubble parameter (top) and the magnitude residuals (bottom) for the Pantheon data compared with the LDE+$H_0$ ML model, relative to the $\Lambda$CDM ML model.  
In the $+H_0$ analysis, the Pantheon absolute magnitude $M$ is marginalized over and so there is an arbitrary constant offset for the data.  For comparison to the $+M$ analysis below, we show the data as offset
from the SH0ES central value by 
\begin{equation}
\label{eq:Mshift}
\Delta M = 5\log_{10} \frac{ 74.03}{67.75} = 0.192.
\end{equation} 

\begin{figure}[t]
\centering
\includegraphics[width=\columnwidth]{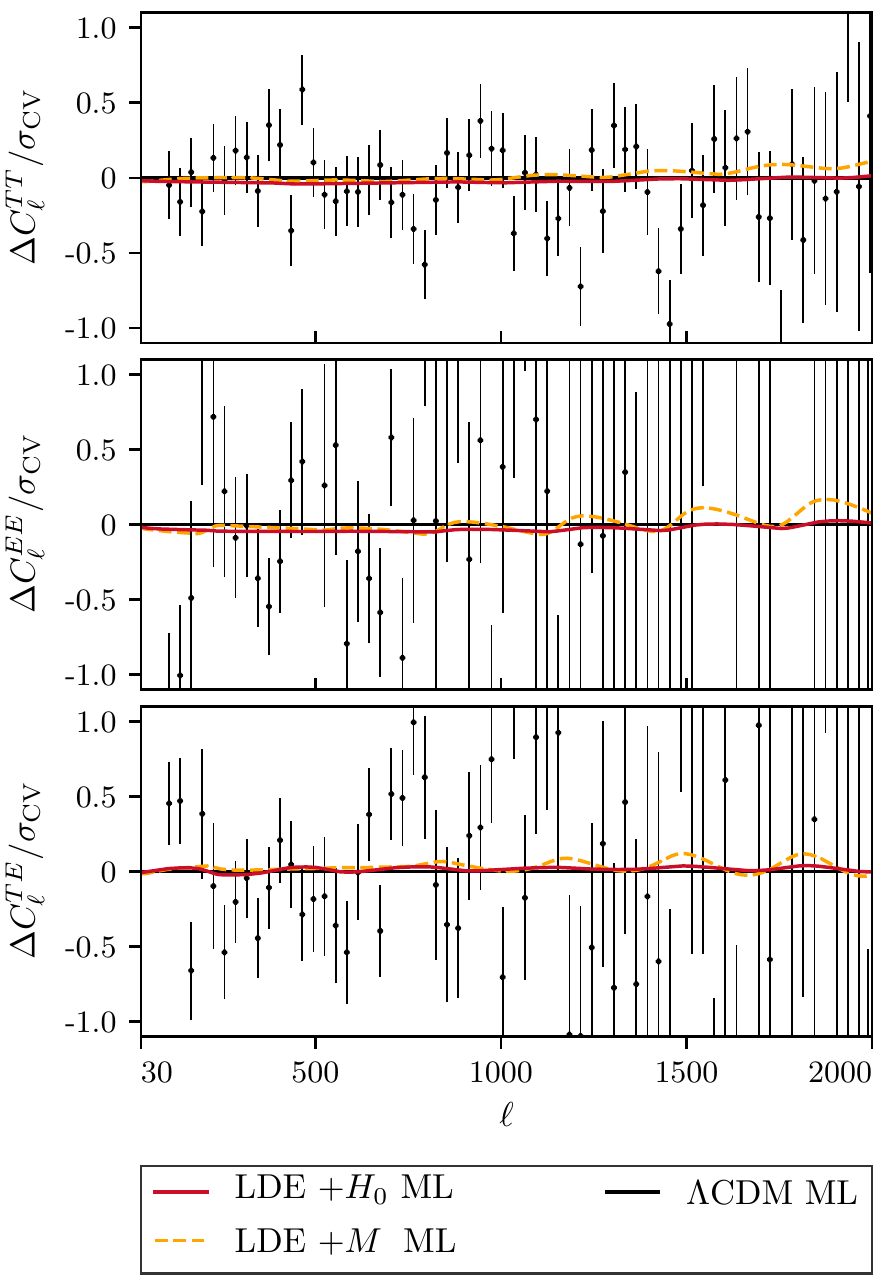}
\caption{
\label{fig:residual}
CMB power spectra residuals with respect to the $\Lambda$CDM ML model  for the Planck data and the LDE  ML model obtained with the $+ H_0$ dataset (red solid) and the
 $+ M$ dataset (orange dashed).}
\end{figure}

  In the $+H_0$ analysis, the LDE+$H_0$ model
would appear to resolve the Hubble tension by changing $H_0$ without significantly distorting the shape
of the Pantheon SN Hubble diagram.   Similarly, we also show the residuals for  the Planck CMB power spectra data in Fig.~\ref{fig:residual} relative to $\Lambda$CDM ML model.
Note that we plot the CMB residuals in units of the cosmic variance per $\ell$ mode, computed as:
\begin{equation}
\sigma_{\rm CV} =
\begin{cases}
\sqrt{\frac{2}{2\ell+1}} C_\ell^{TT}, & TT \,;\\
\sqrt{\frac{1}{2\ell+1}}  \sqrt{ C_\ell^{TT} C_\ell^{EE} + (C_\ell^{TE})^2}, & TE \,;\\
\sqrt{\frac{2}{2\ell+1}}  C_\ell^{EE}, & EE \,. \\
\end{cases}
\end{equation}
The two models are indistinguishable in the CMB even at the cosmic variance limit.
Thus the LDE+$H_0$ model  provides  an equally good fit to the baseline datasets $\Delta \chi^2_{\rm base}\approx 0$, even though $H_0\approx 74$.   Therefore $\Delta \chi^2_{{\rm base}+H_0} \approx -19.5$ when $H_0$ data is included, which satisfies the first
criterion in Eq.~(\ref{eqn:resolution}) for a resolution of the Hubble tension. 
   
\begin{figure}[t]
\centering
\includegraphics[width=\columnwidth]{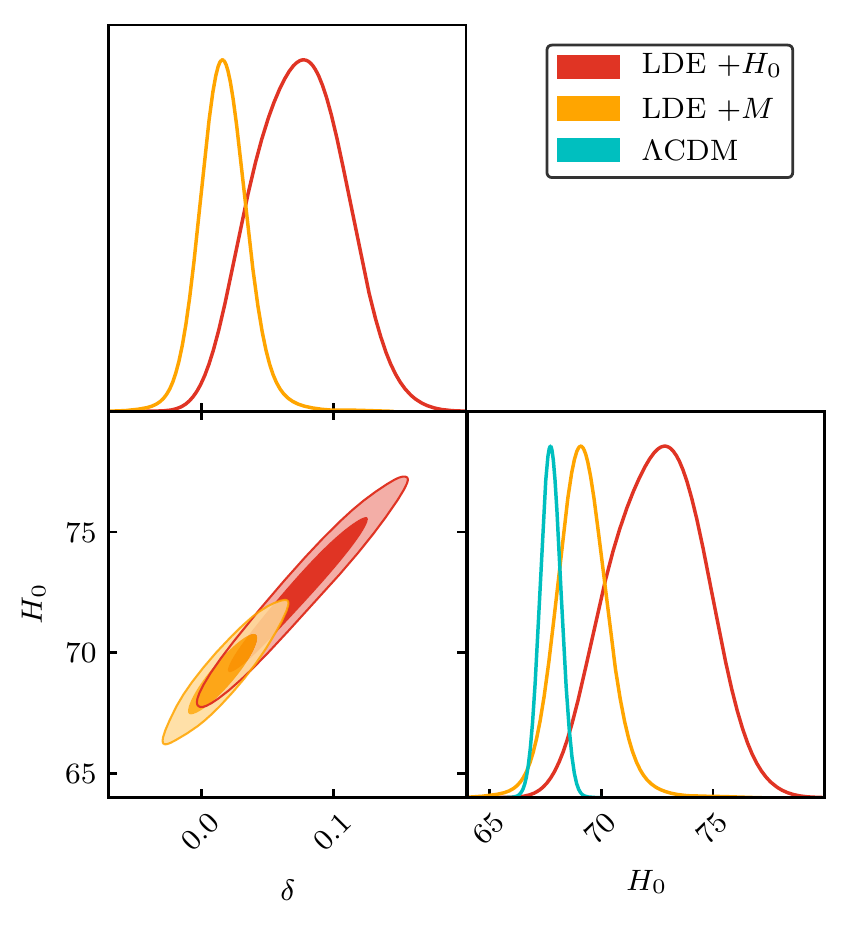}
\caption{
\label{fig:H0}
The marginalized joint posterior for the $\delta$ and $H_0$ parameters in LDE +$H_0$ and LDE +$M$. We also show for comparison the $\Lambda$CDM result for the $H_0$ posterior.
}
\end{figure}

In Fig.~\ref{fig:H0}, we show the posterior constraints on $\delta$ and $H_0$.   
Notice that in the $+H_0$ case, the two are highly correlated and a positive value for $\delta$ is significantly preferred.   
The redshift parameters $z_t,\Delta z$ are on the other hand poorly constrained since any model where the transition occurs well below the last Pantheon SN performs equally well.   
Because of the assumed prior volume in $z_t,\Delta z$ the marginalized constraints on $\delta$ or $H_0$ do not reflect the large improvement of the maximum likelihood model.   
Note that the mean $\delta$ in Tab.~\ref{tab:parameters} is shifted down from the maximum because of the non-Gaussianity of the $z_t,\Delta z$ posterior.

On the other hand, although the LDE model can successfully raise $H_0$ to 74 or beyond, it cannot solve the actual source of $H_0$ tension in the distance ladder.     
When considering the baseline+$M$ analysis,  the LDE+$H_0$ ML model performs
insignificantly better than the $\Lambda$CDM ML, specifically $\Delta \chi^2_{{\rm base}+M}=-0.75$, and so violates the second criterion in Eq.~(\ref{eqn:resolution}).  This is because the model places
the transition below the last SN redshift of the
Pantheon dataset.   Thus $H_0$ is no longer constrained by the SN data in the Hubble flow whereas the
constraint on $M$ still does impact such data.  This is apparent in Fig.~\ref{fig:SN_residual}.   The data there have been offset by $\Delta M = 0.192$ from their SH0ES calibrated central
values, whereas the error in the latter is $\sigma_M =0.042$.

When the LDE model is instead analyzed with the baseline+$M$ data
there is a much weaker improvement in the ML solution 
of $\Delta \chi^2_{{\rm base}+M}=-3.0$.  This model places the transition at $z_t=0.043$ with a much
smaller amplitude of $\delta = 0.015$.  Correspondingly $H_0$ itself only rises to 69.06.
This behavior is in agreement with the result obtained in~\cite{Raveri:2019mxg} using a more general procedure that reconstructs the 
low redshift expansion history under certain prior smoothness constraints.

In Fig.~\ref{fig:SN_residual} we also compare this LDE+$M$ model with the Pantheon data. 
Notice that the transition exploits the slightly lower average magnitudes of the $z \lesssim 0.08$ SN.  On the other hand, the data points are still offset by a much larger $\Delta M=0.192$ from
the central SH0ES calibration. This is reflected in Tab.~\ref{tab:parameters} by the joint
posterior constraint $M=-19.399  \pm 0.012$ which is in tension with the SH0ES calibration
$-19.22 \pm 0.042$ in the same way $H_0$ is in $\Lambda$CDM.  Note also that if we analyze the baseline dataset in this way where $M$ is retained in the Pantheon likelihood
but without the SH0ES constraint itself we obtain $M= -19.415 \pm 0.013$ or $4.4\sigma$ tension with SH0ES, equivalent to the full $H_0$ tension in $\Lambda$CDM, despite marginalizing over late time transitions.
In Fig.~\ref{fig:residual} we show that the Planck residuals for the LDE+$M$ ML model are also slightly higher.   This reflects
the trade-off between  fitting the high redshift data and the distance ladder data and contributes
to the small penalty of $\Delta\chi^2_{\rm base}=0.9$ for the model.  

Note that $\Delta \chi^2_{{\rm base}+M}$  may slightly underestimate the actual improvement
associated with the LDE+$M$ model.   Our technique  assumes that the $\Lambda$CDM shape of the SN Hubble diagram is a good fit to the Pantheon data
when transferring the SH0ES calibration of $M$ to Pantheon.  To the extent that the shape differs and the  redshift weights of the surveys also differ, this transferal depends on the model.   As an upper bound on
the improvement we can take the baseline+$H_0$ analysis of the LDE+$M$ ML model which gives $\Delta \chi^2_{{\rm base}+H_0}=-6.3$.   This would reflect the true improvement if  all of the SH0ES SN used to calibrate $H_0$
are well below the transition in redshift, whereas in reality SN out to $z<0.15$ are employed.   Analogously,
we can think of this as an upper bound on the bias of $\Delta M = (5/\ln 10) \delta \approx 0.03$ on transferring the
calibration, which is within the errors of Eq.~(\ref{eq:Merror}).   Moreover this is much less than the shift of $\Delta M \approx 0.2$  as in Eq.~(\ref{eq:Mshift}) required to fully eliminate the Hubble tension.
It is therefore clear that the LDE model can only  reduce but not fully resolve the
source of $H_0$ tension in the local distance ladder despite being able to raise
$H_0$ to 74 or higher.
The  posterior constraints on $H_0$ and $\delta$ are shown in Fig.~\ref{fig:H0} and
correspond to $H_0 = 69.17\pm 1.09$.  This range  quantifies the ability of LDE to reduce the Hubble tension using $H_0$ values as an effective metric.   Though higher values for $H_0$ itself are clearly still  allowed in the 
sense of an equal goodness of fit, they are strongly disfavored by the small prior volume for $z_t ,\Delta z \approx 0$ and moreover do not reflect a  true resolution of the Hubble tension as our analysis of the inferred absolute magnitude $M$ shows.

\section{Discussion} \label{Sec:Outlook}
We have examined the ability of a late dark energy transition to raise $H_0$ and more generally to solve the highly statistically significant tension between its  local and high redshift determinations in $\Lambda$CDM.   This LDE model is the cosmological analog of a local void scenario for this
tension.   By placing a transition at very low redshift $z \ll 0.1$, such a model can raise
the local expansion rate to $H_0=74$ or beyond without significantly changing any cosmological observables at $z \gg 0.1$, here represented by the CMB temperature, polarization and lensing power spectra, galaxy BAO, and  supernovae relative distances.  
Naively, this would fully resolve the tension and lead to better
fits compared with $\Lambda$CDM+$H_0$ model by $\Delta\chi^2 \sim -18$ for essentially
one extra parameter, the amplitude of the transition $\delta$.  

On the other hand, just as in void scenarios, the problem with such a solution is that  if the transition 
occurs at a redshift $z\lesssim 0.01$, raising $H_0$ does not actually resolve the origin of the Hubble tension which resides in the SH0ES calibrated SN in the Hubble flow, whereas if it occurs at a higher redshift it is constrained to be much smaller than the required amplitude of $\Delta m \sim 0.2$  by the shape of the SN Hubble diagram.    

To properly analyze such cases, we have adopted an alternate approach where the SH0ES local distance ladder calibrates the absolute magnitude of the Pantheon supernovae using 
$\Lambda$CDM as a reference to cross calibrate the samples.  
In this case, even with the flexibility to set the amplitude, location and width of the
transition one can only partially relax the Hubble tension, leading to $H_0=69.06$ at maximum likelihood or constraints of $H_0=69.17 \pm 1.09$.    This relaxation is associated with the
ability to put a  small amplitude $\Delta m \sim 0.03$ step in the Pantheon SN Hubble diagram at $z \sim0.08$ whereas a full resolution to the Hubble tension requires a much larger
change of $\Delta m \sim 0.2$ which is strongly rejected.  While this work was being completed, Ref.~\cite{Dhawan:2020xmp} reached compatible conclusions for the local distance ladder using an unoptimized test transition from Ref.~\cite{Mortonson_2009}.

More generally, our technique should be used  in place of the standard approach of adding
$H_0$ as separate constraint for any dynamical dark energy model that allows features in
the equation of state with $\Delta z\ll 0.1$.  In the future this technique could be improved by a more direct calibration of the effective absolute magnitude of cosmological supernovae datasets.

\acknowledgments
We thank Rick Kessler, Meng-Xiang Lin and Weikang Lin for useful comments.  WH was supported by by U.S.~Dept.~of Energy contract DE-FG02-13ER41958 and the Simons Foundation. 
MR is supported in part by NASA ATP Grant No. NNH17ZDA001N, and by funds provided by the Center for Particle Cosmology. 
Computing resources were provided by the University of Chicago Research Computing Center through the Kavli Institute for Cosmological Physics at the University of Chicago. 

\vfill
\bibliography{biblio}

\begin{thebibliography}{35}%
\makeatletter
\providecommand \@ifxundefined [1]{%
 \@ifx{#1\undefined}
}%
\providecommand \@ifnum [1]{%
 \ifnum #1\expandafter \@firstoftwo
 \else \expandafter \@secondoftwo
 \fi
}%
\providecommand \@ifx [1]{%
 \ifx #1\expandafter \@firstoftwo
 \else \expandafter \@secondoftwo
 \fi
}%
\providecommand \natexlab [1]{#1}%
\providecommand \enquote  [1]{``#1''}%
\providecommand \bibnamefont  [1]{#1}%
\providecommand \bibfnamefont [1]{#1}%
\providecommand \citenamefont [1]{#1}%
\providecommand \href@noop [0]{\@secondoftwo}%
\providecommand \href [0]{\begingroup \@sanitize@url \@href}%
\providecommand \@href[1]{\@@startlink{#1}\@@href}%
\providecommand \@@href[1]{\endgroup#1\@@endlink}%
\providecommand \@sanitize@url [0]{\catcode `\\12\catcode `\$12\catcode
  `\&12\catcode `\#12\catcode `\^12\catcode `\_12\catcode `\%12\relax}%
\providecommand \@@startlink[1]{}%
\providecommand \@@endlink[0]{}%
\providecommand \url  [0]{\begingroup\@sanitize@url \@url }%
\providecommand \@url [1]{\endgroup\@href {#1}{\urlprefix }}%
\providecommand \urlprefix  [0]{URL }%
\providecommand \Eprint [0]{\href }%
\providecommand \doibase [0]{http://dx.doi.org/}%
\providecommand \selectlanguage [0]{\@gobble}%
\providecommand \bibinfo  [0]{\@secondoftwo}%
\providecommand \bibfield  [0]{\@secondoftwo}%
\providecommand \translation [1]{[#1]}%
\providecommand \BibitemOpen [0]{}%
\providecommand \bibitemStop [0]{}%
\providecommand \bibitemNoStop [0]{.\EOS\space}%
\providecommand \EOS [0]{\spacefactor3000\relax}%
\providecommand \BibitemShut  [1]{\csname bibitem#1\endcsname}%
\let\auto@bib@innerbib\@empty
\bibitem [{\citenamefont {Aghanim}\ \emph {et~al.}(2018)\citenamefont {Aghanim}
  \emph {et~al.}}]{Aghanim:2018eyx}%
  \BibitemOpen
  \bibfield  {author} {\bibinfo {author} {\bibfnamefont {N.}~\bibnamefont
  {Aghanim}} \emph {et~al.} (\bibinfo {collaboration} {Planck}),\ }\href@noop
  {} {\  (\bibinfo {year} {2018})},\ \Eprint {http://arxiv.org/abs/1807.06209}
  {arXiv:1807.06209 [astro-ph.CO]} \BibitemShut {NoStop}%
\bibitem [{\citenamefont {Riess}\ \emph {et~al.}(2019)\citenamefont {Riess},
  \citenamefont {Casertano}, \citenamefont {Yuan}, \citenamefont {Macri},\ and\
  \citenamefont {Scolnic}}]{Riess:2019cxk}%
  \BibitemOpen
  \bibfield  {author} {\bibinfo {author} {\bibfnamefont {A.~G.}\ \bibnamefont
  {Riess}}, \bibinfo {author} {\bibfnamefont {S.}~\bibnamefont {Casertano}},
  \bibinfo {author} {\bibfnamefont {W.}~\bibnamefont {Yuan}}, \bibinfo {author}
  {\bibfnamefont {L.~M.}\ \bibnamefont {Macri}}, \ and\ \bibinfo {author}
  {\bibfnamefont {D.}~\bibnamefont {Scolnic}},\ }\href {\doibase
  10.3847/1538-4357/ab1422} {\bibfield  {journal} {\bibinfo  {journal}
  {Astrophys. J.}\ }\textbf {\bibinfo {volume} {876}},\ \bibinfo {pages} {85}
  (\bibinfo {year} {2019})}\BibitemShut {NoStop}%
\bibitem [{\citenamefont {Freedman}\ \emph {et~al.}(2019)\citenamefont
  {Freedman}, \citenamefont {Madore}, \citenamefont {Hatt}, \citenamefont
  {Hoyt}, \citenamefont {Jang}, \citenamefont {Beaton}, \citenamefont {Burns},
  \citenamefont {Lee}, \citenamefont {Monson}, \citenamefont {Neeley},\ and\
  \citenamefont {et~al.}}]{Freedman:2019jwv}%
  \BibitemOpen
  \bibfield  {author} {\bibinfo {author} {\bibfnamefont {W.~L.}\ \bibnamefont
  {Freedman}}, \bibinfo {author} {\bibfnamefont {B.~F.}\ \bibnamefont
  {Madore}}, \bibinfo {author} {\bibfnamefont {D.}~\bibnamefont {Hatt}},
  \bibinfo {author} {\bibfnamefont {T.~J.}\ \bibnamefont {Hoyt}}, \bibinfo
  {author} {\bibfnamefont {I.~S.}\ \bibnamefont {Jang}}, \bibinfo {author}
  {\bibfnamefont {R.~L.}\ \bibnamefont {Beaton}}, \bibinfo {author}
  {\bibfnamefont {C.~R.}\ \bibnamefont {Burns}}, \bibinfo {author}
  {\bibfnamefont {M.~G.}\ \bibnamefont {Lee}}, \bibinfo {author} {\bibfnamefont
  {A.~J.}\ \bibnamefont {Monson}}, \bibinfo {author} {\bibfnamefont {J.~R.}\
  \bibnamefont {Neeley}}, \ and\ \bibinfo {author} {\bibnamefont {et~al.}},\
  }\href {\doibase 10.3847/1538-4357/ab2f73} {\bibfield  {journal} {\bibinfo
  {journal} {Astrophys. J.}\ }\textbf {\bibinfo {volume} {882}},\ \bibinfo
  {pages} {34} (\bibinfo {year} {2019})}\BibitemShut {NoStop}%
\bibitem [{\citenamefont {Freedman}\ \emph {et~al.}(2020)\citenamefont
  {Freedman}, \citenamefont {Madore}, \citenamefont {Hoyt}, \citenamefont
  {Jang}, \citenamefont {Beaton}, \citenamefont {Lee}, \citenamefont {Monson},
  \citenamefont {Neeley},\ and\ \citenamefont {Rich}}]{Freedman:2020dne}%
  \BibitemOpen
  \bibfield  {author} {\bibinfo {author} {\bibfnamefont {W.~L.}\ \bibnamefont
  {Freedman}}, \bibinfo {author} {\bibfnamefont {B.~F.}\ \bibnamefont
  {Madore}}, \bibinfo {author} {\bibfnamefont {T.}~\bibnamefont {Hoyt}},
  \bibinfo {author} {\bibfnamefont {I.~S.}\ \bibnamefont {Jang}}, \bibinfo
  {author} {\bibfnamefont {R.}~\bibnamefont {Beaton}}, \bibinfo {author}
  {\bibfnamefont {M.~G.}\ \bibnamefont {Lee}}, \bibinfo {author} {\bibfnamefont
  {A.}~\bibnamefont {Monson}}, \bibinfo {author} {\bibfnamefont
  {J.}~\bibnamefont {Neeley}}, \ and\ \bibinfo {author} {\bibfnamefont
  {J.}~\bibnamefont {Rich}},\ }\href {\doibase 10.3847/1538-4357/ab7339}
  {\bibfield  {journal} {\bibinfo  {journal} {Astrophys. J.}\ }\textbf
  {\bibinfo {volume} {891}},\ \bibinfo {pages} {57} (\bibinfo {year}
  {2020})}\BibitemShut {NoStop}%
\bibitem [{\citenamefont {Yuan}\ \emph {et~al.}(2019)\citenamefont {Yuan},
  \citenamefont {Riess}, \citenamefont {Macri}, \citenamefont {Casertano},\
  and\ \citenamefont {Scolnic}}]{yuan2019consistent}%
  \BibitemOpen
  \bibfield  {author} {\bibinfo {author} {\bibfnamefont {W.}~\bibnamefont
  {Yuan}}, \bibinfo {author} {\bibfnamefont {A.~G.}\ \bibnamefont {Riess}},
  \bibinfo {author} {\bibfnamefont {L.~M.}\ \bibnamefont {Macri}}, \bibinfo
  {author} {\bibfnamefont {S.}~\bibnamefont {Casertano}}, \ and\ \bibinfo
  {author} {\bibfnamefont {D.}~\bibnamefont {Scolnic}},\ }\href {\doibase
  10.3847/1538-4357/ab4bc9} {\bibfield  {journal} {\bibinfo  {journal}
  {Astrophys. J.}\ }\textbf {\bibinfo {volume} {886}},\ \bibinfo {pages} {61}
  (\bibinfo {year} {2019})}\BibitemShut {NoStop}%
\bibitem [{\citenamefont {Huang}\ \emph {et~al.}(2020)\citenamefont {Huang},
  \citenamefont {Riess}, \citenamefont {Yuan}, \citenamefont {Macri},
  \citenamefont {Zakamska}, \citenamefont {Casertano}, \citenamefont
  {Whitelock}, \citenamefont {Hoffmann}, \citenamefont {Filippenko},\ and\
  \citenamefont {Scolnic}}]{2020ApJ...889....5H}%
  \BibitemOpen
  \bibfield  {author} {\bibinfo {author} {\bibfnamefont {C.~D.}\ \bibnamefont
  {Huang}}, \bibinfo {author} {\bibfnamefont {A.~G.}\ \bibnamefont {Riess}},
  \bibinfo {author} {\bibfnamefont {W.}~\bibnamefont {Yuan}}, \bibinfo {author}
  {\bibfnamefont {L.~M.}\ \bibnamefont {Macri}}, \bibinfo {author}
  {\bibfnamefont {N.~L.}\ \bibnamefont {Zakamska}}, \bibinfo {author}
  {\bibfnamefont {S.}~\bibnamefont {Casertano}}, \bibinfo {author}
  {\bibfnamefont {P.~A.}\ \bibnamefont {Whitelock}}, \bibinfo {author}
  {\bibfnamefont {S.~L.}\ \bibnamefont {Hoffmann}}, \bibinfo {author}
  {\bibfnamefont {A.~V.}\ \bibnamefont {Filippenko}}, \ and\ \bibinfo {author}
  {\bibfnamefont {D.}~\bibnamefont {Scolnic}},\ }\href {\doibase
  10.3847/1538-4357/ab5dbd} {\bibfield  {journal} {\bibinfo  {journal}
  {Astrophys. J.}\ }\textbf {\bibinfo {volume} {889}},\ \bibinfo {pages} {5}
  (\bibinfo {year} {2020})}\BibitemShut {NoStop}%
\bibitem [{\citenamefont {Wong}\ \emph {et~al.}(2019)\citenamefont {Wong} \emph
  {et~al.}}]{Wong:2019kwg}%
  \BibitemOpen
  \bibfield  {author} {\bibinfo {author} {\bibfnamefont {K.~C.}\ \bibnamefont
  {Wong}} \emph {et~al.},\ }\href@noop {} {\  (\bibinfo {year} {2019})},\
  \Eprint {http://arxiv.org/abs/1907.04869} {arXiv:1907.04869 [astro-ph.CO]}
  \BibitemShut {NoStop}%
\bibitem [{\citenamefont {Pandey}\ \emph
  {et~al.}(2019{\natexlab{a}})\citenamefont {Pandey}, \citenamefont {Raveri},\
  and\ \citenamefont {Jain}}]{Pandey:2019yic}%
  \BibitemOpen
  \bibfield  {author} {\bibinfo {author} {\bibfnamefont {S.}~\bibnamefont
  {Pandey}}, \bibinfo {author} {\bibfnamefont {M.}~\bibnamefont {Raveri}}, \
  and\ \bibinfo {author} {\bibfnamefont {B.}~\bibnamefont {Jain}},\ }\href@noop
  {} {\  (\bibinfo {year} {2019}{\natexlab{a}})},\ \Eprint
  {http://arxiv.org/abs/1912.04325} {arXiv:1912.04325 [astro-ph.CO]}
  \BibitemShut {NoStop}%
\bibitem [{\citenamefont {Pesce}\ \emph {et~al.}(2020)\citenamefont {Pesce},
  \citenamefont {Braatz}, \citenamefont {Reid}, \citenamefont {Riess},
  \citenamefont {Scolnic}, \citenamefont {Condon}, \citenamefont {Gao},
  \citenamefont {Henkel}, \citenamefont {Impellizzeri}, \citenamefont {Kuo},\
  and\ \citenamefont {et~al.}}]{Pesce:2020xfe}%
  \BibitemOpen
  \bibfield  {author} {\bibinfo {author} {\bibfnamefont {D.~W.}\ \bibnamefont
  {Pesce}}, \bibinfo {author} {\bibfnamefont {J.~A.}\ \bibnamefont {Braatz}},
  \bibinfo {author} {\bibfnamefont {M.~J.}\ \bibnamefont {Reid}}, \bibinfo
  {author} {\bibfnamefont {A.~G.}\ \bibnamefont {Riess}}, \bibinfo {author}
  {\bibfnamefont {D.}~\bibnamefont {Scolnic}}, \bibinfo {author} {\bibfnamefont
  {J.~J.}\ \bibnamefont {Condon}}, \bibinfo {author} {\bibfnamefont
  {F.}~\bibnamefont {Gao}}, \bibinfo {author} {\bibfnamefont {C.}~\bibnamefont
  {Henkel}}, \bibinfo {author} {\bibfnamefont {C.~M.~V.}\ \bibnamefont
  {Impellizzeri}}, \bibinfo {author} {\bibfnamefont {C.~Y.}\ \bibnamefont
  {Kuo}}, \ and\ \bibinfo {author} {\bibnamefont {et~al.}},\ }\href {\doibase
  10.3847/2041-8213/ab75f0} {\bibfield  {journal} {\bibinfo  {journal}
  {Astrophys. J. Lett.}\ }\textbf {\bibinfo {volume} {891}},\ \bibinfo {pages}
  {L1} (\bibinfo {year} {2020})}\BibitemShut {NoStop}%
\bibitem [{\citenamefont {Bernal}\ \emph {et~al.}(2016)\citenamefont {Bernal},
  \citenamefont {Verde},\ and\ \citenamefont {Riess}}]{Bernal:2016gxb}%
  \BibitemOpen
  \bibfield  {author} {\bibinfo {author} {\bibfnamefont {J.~L.}\ \bibnamefont
  {Bernal}}, \bibinfo {author} {\bibfnamefont {L.}~\bibnamefont {Verde}}, \
  and\ \bibinfo {author} {\bibfnamefont {A.~G.}\ \bibnamefont {Riess}},\ }\href
  {\doibase 10.1088/1475-7516/2016/10/019} {\bibfield  {journal} {\bibinfo
  {journal} {J. Cosmol. Astropart. Phys.}\ }\textbf {\bibinfo {volume}
  {2016}},\ \bibinfo {pages} {019–019} (\bibinfo {year} {2016})}\BibitemShut
  {NoStop}%
\bibitem [{\citenamefont {Aylor}\ \emph {et~al.}(2019)\citenamefont {Aylor},
  \citenamefont {Joy}, \citenamefont {Knox}, \citenamefont {Millea},
  \citenamefont {Raghunathan},\ and\ \citenamefont {Wu}}]{Aylor:2018drw}%
  \BibitemOpen
  \bibfield  {author} {\bibinfo {author} {\bibfnamefont {K.}~\bibnamefont
  {Aylor}}, \bibinfo {author} {\bibfnamefont {M.}~\bibnamefont {Joy}}, \bibinfo
  {author} {\bibfnamefont {L.}~\bibnamefont {Knox}}, \bibinfo {author}
  {\bibfnamefont {M.}~\bibnamefont {Millea}}, \bibinfo {author} {\bibfnamefont
  {S.}~\bibnamefont {Raghunathan}}, \ and\ \bibinfo {author} {\bibfnamefont
  {W.~L.~K.}\ \bibnamefont {Wu}},\ }\href {\doibase 10.3847/1538-4357/ab0898}
  {\bibfield  {journal} {\bibinfo  {journal} {Astrophys. J.}\ }\textbf
  {\bibinfo {volume} {874}},\ \bibinfo {pages} {4} (\bibinfo {year}
  {2019})}\BibitemShut {NoStop}%
\bibitem [{\citenamefont {Knox}\ and\ \citenamefont
  {Millea}(2020)}]{Knox:2019rjx}%
  \BibitemOpen
  \bibfield  {author} {\bibinfo {author} {\bibfnamefont {L.}~\bibnamefont
  {Knox}}\ and\ \bibinfo {author} {\bibfnamefont {M.}~\bibnamefont {Millea}},\
  }\href {\doibase 10.1103/physrevd.101.043533} {\bibfield  {journal} {\bibinfo
   {journal} {Phys. Rev. D}\ }\textbf {\bibinfo {volume} {101}} (\bibinfo
  {year} {2020}),\ 10.1103/physrevd.101.043533}\BibitemShut {NoStop}%
\bibitem [{\citenamefont {Poulin}\ \emph {et~al.}(2019)\citenamefont {Poulin},
  \citenamefont {Smith}, \citenamefont {Karwal},\ and\ \citenamefont
  {Kamionkowski}}]{Poulin:2018cxd}%
  \BibitemOpen
  \bibfield  {author} {\bibinfo {author} {\bibfnamefont {V.}~\bibnamefont
  {Poulin}}, \bibinfo {author} {\bibfnamefont {T.~L.}\ \bibnamefont {Smith}},
  \bibinfo {author} {\bibfnamefont {T.}~\bibnamefont {Karwal}}, \ and\ \bibinfo
  {author} {\bibfnamefont {M.}~\bibnamefont {Kamionkowski}},\ }\href {\doibase
  10.1103/PhysRevLett.122.221301} {\bibfield  {journal} {\bibinfo  {journal}
  {Phys. Rev. Lett.}\ }\textbf {\bibinfo {volume} {122}},\ \bibinfo {pages}
  {221301} (\bibinfo {year} {2019})}\BibitemShut {NoStop}%
\bibitem [{\citenamefont {Agrawal}\ \emph {et~al.}(2019)\citenamefont
  {Agrawal}, \citenamefont {Cyr-Racine}, \citenamefont {Pinner},\ and\
  \citenamefont {Randall}}]{Agrawal:2019lmo}%
  \BibitemOpen
  \bibfield  {author} {\bibinfo {author} {\bibfnamefont {P.}~\bibnamefont
  {Agrawal}}, \bibinfo {author} {\bibfnamefont {F.-Y.}\ \bibnamefont
  {Cyr-Racine}}, \bibinfo {author} {\bibfnamefont {D.}~\bibnamefont {Pinner}},
  \ and\ \bibinfo {author} {\bibfnamefont {L.}~\bibnamefont {Randall}},\
  }\href@noop {} {\  (\bibinfo {year} {2019})},\ \Eprint
  {http://arxiv.org/abs/1904.01016} {arXiv:1904.01016 [astro-ph.CO]}
  \BibitemShut {NoStop}%
\bibitem [{\citenamefont {Lin}\ \emph {et~al.}(2019)\citenamefont {Lin},
  \citenamefont {Benevento}, \citenamefont {Hu},\ and\ \citenamefont
  {Raveri}}]{Lin:2019qug}%
  \BibitemOpen
  \bibfield  {author} {\bibinfo {author} {\bibfnamefont {M.-X.}\ \bibnamefont
  {Lin}}, \bibinfo {author} {\bibfnamefont {G.}~\bibnamefont {Benevento}},
  \bibinfo {author} {\bibfnamefont {W.}~\bibnamefont {Hu}}, \ and\ \bibinfo
  {author} {\bibfnamefont {M.}~\bibnamefont {Raveri}},\ }\href {\doibase
  10.1103/PhysRevD.100.063542} {\bibfield  {journal} {\bibinfo  {journal}
  {Phys. Rev. D}\ }\textbf {\bibinfo {volume} {100}},\ \bibinfo {pages}
  {063542} (\bibinfo {year} {2019})}\BibitemShut {NoStop}%
\bibitem [{\citenamefont {Mortonson}\ \emph {et~al.}(2009)\citenamefont
  {Mortonson}, \citenamefont {Hu},\ and\ \citenamefont
  {Huterer}}]{Mortonson_2009}%
  \BibitemOpen
  \bibfield  {author} {\bibinfo {author} {\bibfnamefont {M.}~\bibnamefont
  {Mortonson}}, \bibinfo {author} {\bibfnamefont {W.}~\bibnamefont {Hu}}, \
  and\ \bibinfo {author} {\bibfnamefont {D.}~\bibnamefont {Huterer}},\ }\href
  {\doibase 10.1103/physrevd.80.067301} {\bibfield  {journal} {\bibinfo
  {journal} {Phys. Rev. D}\ }\textbf {\bibinfo {volume} {80}} (\bibinfo {year}
  {2009}),\ 10.1103/physrevd.80.067301}\BibitemShut {NoStop}%
\bibitem [{\citenamefont {Raveri}\ \emph {et~al.}(2019)\citenamefont {Raveri},
  \citenamefont {Zacharegkas},\ and\ \citenamefont
  {Hu}}]{raveri2019quantifying}%
  \BibitemOpen
  \bibfield  {author} {\bibinfo {author} {\bibfnamefont {M.}~\bibnamefont
  {Raveri}}, \bibinfo {author} {\bibfnamefont {G.}~\bibnamefont {Zacharegkas}},
  \ and\ \bibinfo {author} {\bibfnamefont {W.}~\bibnamefont {Hu}},\ }\href@noop
  {} {\  (\bibinfo {year} {2019})},\ \Eprint {http://arxiv.org/abs/1912.04880}
  {arXiv:1912.04880 [astro-ph.CO]} \BibitemShut {NoStop}%
\bibitem [{\citenamefont {Scolnic}\ \emph {et~al.}(2018)\citenamefont {Scolnic}
  \emph {et~al.}}]{Scolnic:2017caz}%
  \BibitemOpen
  \bibfield  {author} {\bibinfo {author} {\bibfnamefont {D.~M.}\ \bibnamefont
  {Scolnic}} \emph {et~al.},\ }\href {\doibase 10.3847/1538-4357/aab9bb}
  {\bibfield  {journal} {\bibinfo  {journal} {Astrophys. J.}\ }\textbf
  {\bibinfo {volume} {859}},\ \bibinfo {pages} {101} (\bibinfo {year}
  {2018})}\BibitemShut {NoStop}%
\bibitem [{\citenamefont {Hu}\ \emph {et~al.}(2014{\natexlab{a}})\citenamefont
  {Hu}, \citenamefont {Raveri}, \citenamefont {Frusciante},\ and\ \citenamefont
  {Silvestri}}]{Hu:2013twa}%
  \BibitemOpen
  \bibfield  {author} {\bibinfo {author} {\bibfnamefont {B.}~\bibnamefont
  {Hu}}, \bibinfo {author} {\bibfnamefont {M.}~\bibnamefont {Raveri}}, \bibinfo
  {author} {\bibfnamefont {N.}~\bibnamefont {Frusciante}}, \ and\ \bibinfo
  {author} {\bibfnamefont {A.}~\bibnamefont {Silvestri}},\ }\href {\doibase
  10.1103/PhysRevD.89.103530} {\bibfield  {journal} {\bibinfo  {journal} {Phys.
  Rev. D}\ }\textbf {\bibinfo {volume} {89}},\ \bibinfo {pages} {103530}
  (\bibinfo {year} {2014}{\natexlab{a}})}\BibitemShut {NoStop}%
\bibitem [{\citenamefont {Raveri}\ \emph {et~al.}(2014)\citenamefont {Raveri},
  \citenamefont {Hu}, \citenamefont {Frusciante},\ and\ \citenamefont
  {Silvestri}}]{Raveri:2014cka}%
  \BibitemOpen
  \bibfield  {author} {\bibinfo {author} {\bibfnamefont {M.}~\bibnamefont
  {Raveri}}, \bibinfo {author} {\bibfnamefont {B.}~\bibnamefont {Hu}}, \bibinfo
  {author} {\bibfnamefont {N.}~\bibnamefont {Frusciante}}, \ and\ \bibinfo
  {author} {\bibfnamefont {A.}~\bibnamefont {Silvestri}},\ }\href {\doibase
  10.1103/PhysRevD.90.043513} {\bibfield  {journal} {\bibinfo  {journal} {Phys.
  Rev. D}\ }\textbf {\bibinfo {volume} {90}},\ \bibinfo {pages} {043513}
  (\bibinfo {year} {2014})}\BibitemShut {NoStop}%
\bibitem [{\citenamefont {Hu}\ \emph {et~al.}(2014{\natexlab{b}})\citenamefont
  {Hu}, \citenamefont {Raveri}, \citenamefont {Frusciante},\ and\ \citenamefont
  {Silvestri}}]{hu2014eftcambeftcosmomc}%
  \BibitemOpen
  \bibfield  {author} {\bibinfo {author} {\bibfnamefont {B.}~\bibnamefont
  {Hu}}, \bibinfo {author} {\bibfnamefont {M.}~\bibnamefont {Raveri}}, \bibinfo
  {author} {\bibfnamefont {N.}~\bibnamefont {Frusciante}}, \ and\ \bibinfo
  {author} {\bibfnamefont {A.}~\bibnamefont {Silvestri}},\ }\href@noop {} {\
  (\bibinfo {year} {2014}{\natexlab{b}})},\ \Eprint
  {http://arxiv.org/abs/1405.3590} {arXiv:1405.3590 [astro-ph.IM]} \BibitemShut
  {NoStop}%
\bibitem [{\citenamefont {Lewis}\ \emph {et~al.}(2000)\citenamefont {Lewis},
  \citenamefont {Challinor},\ and\ \citenamefont {Lasenby}}]{Lewis:1999bs}%
  \BibitemOpen
  \bibfield  {author} {\bibinfo {author} {\bibfnamefont {A.}~\bibnamefont
  {Lewis}}, \bibinfo {author} {\bibfnamefont {A.}~\bibnamefont {Challinor}}, \
  and\ \bibinfo {author} {\bibfnamefont {A.}~\bibnamefont {Lasenby}},\ }\href
  {\doibase 10.1086/309179} {\bibfield  {journal} {\bibinfo  {journal}
  {Astrophys. J.}\ }\textbf {\bibinfo {volume} {538}},\ \bibinfo {pages} {473}
  (\bibinfo {year} {2000})}\BibitemShut {NoStop}%
\bibitem [{\citenamefont {{Gubitosi}}\ \emph {et~al.}(2013)\citenamefont
  {{Gubitosi}}, \citenamefont {{Piazza}},\ and\ \citenamefont
  {{Vernizzi}}}]{2013JCAP...02..032G}%
  \BibitemOpen
  \bibfield  {author} {\bibinfo {author} {\bibfnamefont {G.}~\bibnamefont
  {{Gubitosi}}}, \bibinfo {author} {\bibfnamefont {F.}~\bibnamefont
  {{Piazza}}}, \ and\ \bibinfo {author} {\bibfnamefont {F.}~\bibnamefont
  {{Vernizzi}}},\ }\href {\doibase 10.1088/1475-7516/2013/02/032} {\bibfield
  {journal} {\bibinfo  {journal} {J. Cosmol. Astropart. Phys.}\ }\textbf
  {\bibinfo {volume} {2}},\ \bibinfo {eid} {032} (\bibinfo {year}
  {2013})}\BibitemShut {NoStop}%
\bibitem [{\citenamefont {{Gleyzes}}\ \emph {et~al.}(2013)\citenamefont
  {{Gleyzes}}, \citenamefont {{Langlois}}, \citenamefont {{Piazza}},\ and\
  \citenamefont {{Vernizzi}}}]{2013JCAP...08..025G}%
  \BibitemOpen
  \bibfield  {author} {\bibinfo {author} {\bibfnamefont {J.}~\bibnamefont
  {{Gleyzes}}}, \bibinfo {author} {\bibfnamefont {D.}~\bibnamefont
  {{Langlois}}}, \bibinfo {author} {\bibfnamefont {F.}~\bibnamefont
  {{Piazza}}}, \ and\ \bibinfo {author} {\bibfnamefont {F.}~\bibnamefont
  {{Vernizzi}}},\ }\href {\doibase 10.1088/1475-7516/2013/08/025} {\bibfield
  {journal} {\bibinfo  {journal} {J. Cosmol. Astropart. Phys.}\ }\textbf
  {\bibinfo {volume} {8}},\ \bibinfo {eid} {025} (\bibinfo {year}
  {2013})}\BibitemShut {NoStop}%
\bibitem [{\citenamefont {Lewis}\ and\ \citenamefont
  {Bridle}(2002)}]{Lewis:2002ah}%
  \BibitemOpen
  \bibfield  {author} {\bibinfo {author} {\bibfnamefont {A.}~\bibnamefont
  {Lewis}}\ and\ \bibinfo {author} {\bibfnamefont {S.}~\bibnamefont {Bridle}},\
  }\href {\doibase 10.1103/PhysRevD.66.103511} {\bibfield  {journal} {\bibinfo
  {journal} {Phys. Rev. D}\ }\textbf {\bibinfo {volume} {66}},\ \bibinfo
  {pages} {103511} (\bibinfo {year} {2002})}\BibitemShut {NoStop}%
\bibitem [{\citenamefont {Lewis}(2019)}]{Lewis:2019xzd}%
  \BibitemOpen
  \bibfield  {author} {\bibinfo {author} {\bibfnamefont {A.}~\bibnamefont
  {Lewis}},\ }\href@noop {} {\  (\bibinfo {year} {2019})},\ \Eprint
  {http://arxiv.org/abs/1910.13970} {arXiv:1910.13970 [astro-ph.IM]}
  \BibitemShut {NoStop}%
\bibitem [{\citenamefont {Aghanim}\ \emph {et~al.}(2019)\citenamefont {Aghanim}
  \emph {et~al.}}]{Aghanim:2019ame}%
  \BibitemOpen
  \bibfield  {author} {\bibinfo {author} {\bibfnamefont {N.}~\bibnamefont
  {Aghanim}} \emph {et~al.} (\bibinfo {collaboration} {Planck}),\ }\href@noop
  {} {\  (\bibinfo {year} {2019})},\ \Eprint {http://arxiv.org/abs/1907.12875}
  {arXiv:1907.12875 [astro-ph.CO]} \BibitemShut {NoStop}%
\bibitem [{\citenamefont {Alam}\ \emph {et~al.}(2017)\citenamefont {Alam} \emph
  {et~al.}}]{Alam:2016hwk}%
  \BibitemOpen
  \bibfield  {author} {\bibinfo {author} {\bibfnamefont {S.}~\bibnamefont
  {Alam}} \emph {et~al.} (\bibinfo {collaboration} {BOSS}),\ }\href {\doibase
  10.1093/mnras/stx721} {\bibfield  {journal} {\bibinfo  {journal} {Mon. Not.
  Roy. Astron. Soc.}\ }\textbf {\bibinfo {volume} {470}},\ \bibinfo {pages}
  {2617} (\bibinfo {year} {2017})}\BibitemShut {NoStop}%
\bibitem [{\citenamefont {Ross}\ \emph {et~al.}(2015)\citenamefont {Ross},
  \citenamefont {Samushia}, \citenamefont {Howlett}, \citenamefont {Percival},
  \citenamefont {Burden},\ and\ \citenamefont {Manera}}]{Ross:2014qpa}%
  \BibitemOpen
  \bibfield  {author} {\bibinfo {author} {\bibfnamefont {A.~J.}\ \bibnamefont
  {Ross}}, \bibinfo {author} {\bibfnamefont {L.}~\bibnamefont {Samushia}},
  \bibinfo {author} {\bibfnamefont {C.}~\bibnamefont {Howlett}}, \bibinfo
  {author} {\bibfnamefont {W.~J.}\ \bibnamefont {Percival}}, \bibinfo {author}
  {\bibfnamefont {A.}~\bibnamefont {Burden}}, \ and\ \bibinfo {author}
  {\bibfnamefont {M.}~\bibnamefont {Manera}},\ }\href {\doibase
  10.1093/mnras/stv154} {\bibfield  {journal} {\bibinfo  {journal} {Mon. Not.
  Roy. Astron. Soc.}\ }\textbf {\bibinfo {volume} {449}},\ \bibinfo {pages}
  {835} (\bibinfo {year} {2015})}\BibitemShut {NoStop}%
\bibitem [{\citenamefont {Beutler}\ \emph {et~al.}(2011)\citenamefont
  {Beutler}, \citenamefont {Blake}, \citenamefont {Colless}, \citenamefont
  {Jones}, \citenamefont {Staveley-Smith}, \citenamefont {Campbell},
  \citenamefont {Parker}, \citenamefont {Saunders},\ and\ \citenamefont
  {Watson}}]{Beutler:2011hx}%
  \BibitemOpen
  \bibfield  {author} {\bibinfo {author} {\bibfnamefont {F.}~\bibnamefont
  {Beutler}}, \bibinfo {author} {\bibfnamefont {C.}~\bibnamefont {Blake}},
  \bibinfo {author} {\bibfnamefont {M.}~\bibnamefont {Colless}}, \bibinfo
  {author} {\bibfnamefont {D.~H.}\ \bibnamefont {Jones}}, \bibinfo {author}
  {\bibfnamefont {L.}~\bibnamefont {Staveley-Smith}}, \bibinfo {author}
  {\bibfnamefont {L.}~\bibnamefont {Campbell}}, \bibinfo {author}
  {\bibfnamefont {Q.}~\bibnamefont {Parker}}, \bibinfo {author} {\bibfnamefont
  {W.}~\bibnamefont {Saunders}}, \ and\ \bibinfo {author} {\bibfnamefont
  {F.}~\bibnamefont {Watson}},\ }\href {\doibase
  10.1111/j.1365-2966.2011.19250.x} {\bibfield  {journal} {\bibinfo  {journal}
  {Mon. Not. Roy. Astron. Soc.}\ }\textbf {\bibinfo {volume} {416}},\ \bibinfo
  {pages} {3017} (\bibinfo {year} {2011})}\BibitemShut {NoStop}%
\bibitem [{\citenamefont {Guy}\ \emph {et~al.}(2005)\citenamefont {Guy},
  \citenamefont {Astier}, \citenamefont {Nobili}, \citenamefont {Regnault},\
  and\ \citenamefont {Pain}}]{Guy:2005me}%
  \BibitemOpen
  \bibfield  {author} {\bibinfo {author} {\bibfnamefont {J.}~\bibnamefont
  {Guy}}, \bibinfo {author} {\bibfnamefont {P.}~\bibnamefont {Astier}},
  \bibinfo {author} {\bibfnamefont {S.}~\bibnamefont {Nobili}}, \bibinfo
  {author} {\bibfnamefont {N.}~\bibnamefont {Regnault}}, \ and\ \bibinfo
  {author} {\bibfnamefont {R.}~\bibnamefont {Pain}} (\bibinfo {collaboration}
  {SNLS}),\ }\href {\doibase 10.1051/0004-6361:20053025} {\bibfield  {journal}
  {\bibinfo  {journal} {Astron. Astrophys.}\ }\textbf {\bibinfo {volume}
  {443}},\ \bibinfo {pages} {781} (\bibinfo {year} {2005})}\BibitemShut
  {NoStop}%
\bibitem [{\citenamefont {Pandey}\ \emph
  {et~al.}(2019{\natexlab{b}})\citenamefont {Pandey}, \citenamefont {Raveri},\
  and\ \citenamefont {Jain}}]{p2019model}%
  \BibitemOpen
  \bibfield  {author} {\bibinfo {author} {\bibfnamefont {S.}~\bibnamefont
  {Pandey}}, \bibinfo {author} {\bibfnamefont {M.}~\bibnamefont {Raveri}}, \
  and\ \bibinfo {author} {\bibfnamefont {B.}~\bibnamefont {Jain}},\ }\href@noop
  {} {\  (\bibinfo {year} {2019}{\natexlab{b}})},\ \Eprint
  {http://arxiv.org/abs/1912.04325} {arXiv:1912.04325 [astro-ph.CO]}
  \BibitemShut {NoStop}%
\bibitem [{\citenamefont {Camarena}\ and\ \citenamefont
  {Marra}(2020)}]{Camarena_2020}%
  \BibitemOpen
  \bibfield  {author} {\bibinfo {author} {\bibfnamefont {D.}~\bibnamefont
  {Camarena}}\ and\ \bibinfo {author} {\bibfnamefont {V.}~\bibnamefont
  {Marra}},\ }\href {\doibase 10.1103/physrevresearch.2.013028} {\bibfield
  {journal} {\bibinfo  {journal} {Phys. Rev. Res.}\ }\textbf {\bibinfo {volume}
  {2}} (\bibinfo {year} {2020}),\ 10.1103/physrevresearch.2.013028}\BibitemShut
  {NoStop}%
\bibitem [{\citenamefont {Raveri}(2020)}]{Raveri:2019mxg}%
  \BibitemOpen
  \bibfield  {author} {\bibinfo {author} {\bibfnamefont {M.}~\bibnamefont
  {Raveri}},\ }\href {\doibase 10.1103/physrevd.101.083524} {\bibfield
  {journal} {\bibinfo  {journal} {Phys. Rev. D}\ }\textbf {\bibinfo {volume}
  {101}} (\bibinfo {year} {2020}),\ 10.1103/physrevd.101.083524}\BibitemShut
  {NoStop}%
\bibitem [{\citenamefont {Dhawan}\ \emph {et~al.}(2020)\citenamefont {Dhawan},
  \citenamefont {Brout}, \citenamefont {Scolnic}, \citenamefont {Goobar},
  \citenamefont {Riess},\ and\ \citenamefont {Miranda}}]{Dhawan:2020xmp}%
  \BibitemOpen
  \bibfield  {author} {\bibinfo {author} {\bibfnamefont {S.}~\bibnamefont
  {Dhawan}}, \bibinfo {author} {\bibfnamefont {D.}~\bibnamefont {Brout}},
  \bibinfo {author} {\bibfnamefont {D.}~\bibnamefont {Scolnic}}, \bibinfo
  {author} {\bibfnamefont {A.}~\bibnamefont {Goobar}}, \bibinfo {author}
  {\bibfnamefont {A.~G.}\ \bibnamefont {Riess}}, \ and\ \bibinfo {author}
  {\bibfnamefont {V.}~\bibnamefont {Miranda}},\ }\href@noop {} {\  (\bibinfo
  {year} {2020})},\ \Eprint {http://arxiv.org/abs/2001.09260} {arXiv:2001.09260
  [astro-ph.CO]} \BibitemShut {NoStop}%
\end{thebibliography}%
\end{document}